\documentclass[aps,showpacs,prx,amssymb,amsmath,twocolumn]{revtex4-2}
\usepackage{graphicx}
\usepackage{amssymb}
\usepackage{amsmath}
\usepackage{amsthm}
\usepackage{bm}
\usepackage{xcolor} 
\usepackage{array}
\usepackage{multirow}
\usepackage{tabularx}
\usepackage{booktabs}
\usepackage{textcomp}
\usepackage{mathtools}
\usepackage{hyperref}
\usepackage{qcircuit}
\usepackage{bbm}
\usepackage{cancel}
\usepackage{comment}
\usepackage{physics}

\begin{document}

\title{A Rydberg-atom approach to the integer factorization problem}
\author{Juyoung Park$^1$, Seokho Jeong$^1$, Minhyuk Kim$^{1,2}$, Kangheun Kim$^1$, Andrew Byun$^1$, Louis Vignoli$^3$, Louis-Paul Henry$^3$, Loïc  Henriet$^3$, and Jaewook Ahn$^1$}
\address{$^1$Department of Physics, KAIST, Daejeon 34141, Republic of Korea} 
\address{$^2$Department of Physics, Korea University, Seoul 02841, Republic of Korea} 
\address{$^3$PASQAL, 7 rue Léonard de Vinci, 91300 Massy, France} 
\date{\today}

\begin{abstract} \noindent
The task of factoring integers poses a significant challenge in modern cryptography, and quantum computing holds the potential to efficiently address this problem compared to classical algorithms. Thus, it is crucial to develop quantum computing algorithms to address this problem. This study introduces a quantum approach that utilizes Rydberg atoms to tackle the factorization problem. Experimental demonstrations are conducted for the factorization of small composite numbers such as $6 = 2 \times 3$, $15 = 3 \times 5$, and $35 = 5 \times 7$. This approach involves employing Rydberg-atom graphs to algorithmically program binary multiplication tables, yielding many-body ground states that represent superpositions of factoring solutions. Subsequently, these states are probed using quantum adiabatic computing. Limitations of this method are discussed, specifically addressing the scalability of current Rydberg quantum computing for the intricate computational problem. 
\end{abstract}

\maketitle

\section{Introduction} 
Modern cryptosystems using public-key distribution rely on the fact that finding prime factors, $p$ and $q$, of a given semi-prime integer, $n=p\times q$, is computationally inefficient in classical computation~\cite{Rivest1978RSA}. On a quantum computer, Shor's algorithm is expected to run in a poly-logarithimic time of $n$, i.e., to solve the factorization problem efficiently~\cite{Shor1994,Beckman1996}. Experimental tests of Shor's algorithm for small integers have been conducted on various quantum gate-based computers, including those using  NMR~\cite{Vandersypen2001}, trapped ions~\cite{Monz2016}, superconductor qubits~\cite{Skosana2021,Yan2022}, and photons~\cite{Lu2007,Lanyon2007}.  Improvements are expected in gate fidelity and system size to facilitate factorization of larger numbers. An alternative approach is provided by quantum adiabatic computing~\cite{Albash2018}, 
where the integer factorization problem is encoded into the Hamiltonian of a quantum many-body system, which allows the prime factors to be obtained by adiabatically driving the system to its ground state. Scalable experiments of quantum adiabatic methods have been carried out with NMR systems~\cite{Peng2008,Xu2012} and on commercially accessible platforms such as IBMQ~\cite{Saxena2021} and D-Wave~\cite{Jiang2018}. These experiments utilize quadratic-unconstrained-binary-optimization (QUBO) to encode the factorization problem into Hamiltonians.

In recent years, there has been a rapid progress in the field of Rydberg-atom based quantum computing~\cite{Saffman2010,Wu2021,Morgado2021,Minhyuk2023}. Atomic qubits numbering in the hundreds have become available and are used for various quantum applications, including quantum simulations, adiabatic quantum computing, and quantum approximate optimization algorithms~\cite{Ebadi2021,Scholl2021,Semeghini2021,Chen2023,Bluvstein2022,Graham2022,Bluvstein2023}. The adaptability of atom rearrangement methods has enabled the creation of nearly arbitrary atomic graphs, $G(V,E)$, where $V$ and $E$ represent of atoms and pair-wise strong Rydberg couplings, respectively~\cite{Minhyuk2023}. These resulting Rydberg-atom graphs are applied to nondeterministic polynomial time (NP)-complete problems including the maximum independent set (MIS), maximum cut, and satisfiability (SAT) problems~\cite{EbadiScience2022_postprocessing,Minhyuk2022,Byun2022,Graham2022,Jeong2023}.

This paper aims to utilize Rydberg-atom graphs to program the integer factorization problem and to determine the integer factors experimentally. The procedure involves an efficient two-step reduction algorithm that transforms the integer factorization problem, via (i) the SAT problem, into (ii) the MIS problem. Additionally, it includes a protocol for embedding the MIS problem onto a Rydberg-atom graph and experimentally probing the Rydberg-atom graph's ground state. The subsequent sections of the paper elaborate on the procedure, starting with an overview of the use of a Rydberg-atom graph in integer factorization, illustrated by the  example of $p\times q=6$ in Sec.~\ref{SecII}. Details about encoding of the factorization into the SAT problem are provided in Sec.~\ref{SecIII}. Experimental demonstrations of $p\times q=15$ and $35$ are presented in Sec.~\ref{SecIV}. Scalability issues related to the Rydberg-atom approach to the integer factorization problem are discussed in Sec.~\ref{SecV}, leading to conclusions in Sec.~\ref{SecVI}.

\section{Rydberg-atom approach to the integer factorization problem} \label{SecII} 

\begin{figure*}[htbp]
  \centering
\includegraphics[width=1\textwidth]{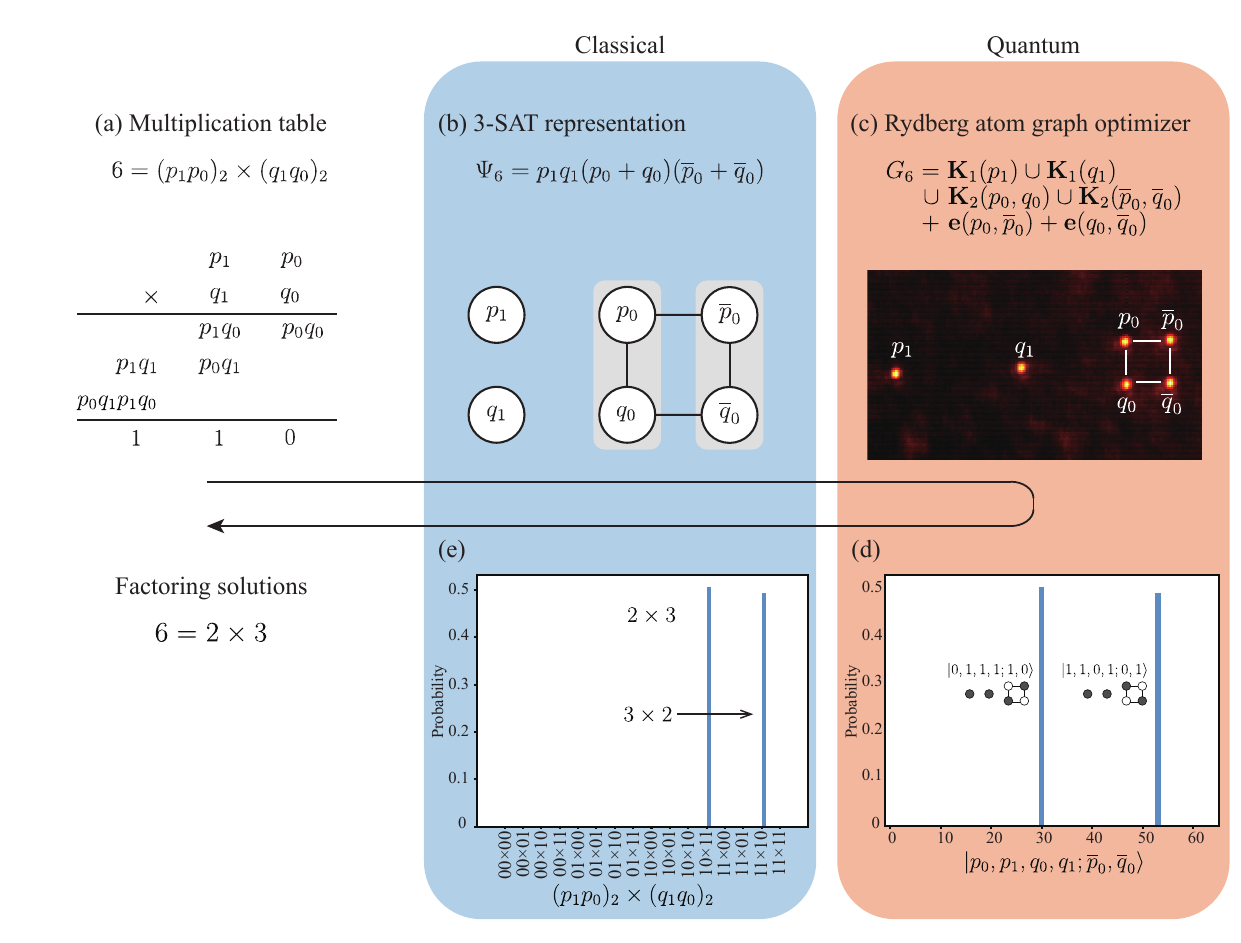}
\caption{The procedure of addressing the integer factorization problem using a Rydberg-atom graph. (a) The instance $p\times q=6$ and its associated multiplication table. (b) The 3-satisfiability (3-SAT) representation of the given factoring problem, along with the corresponding graph $G_6$. (c) Implementation of Rydberg atom quantum adiabatic computing, where $G_6$ is manifested as a Rydberg-atom graph. (d,e) Experimental probability distributions in $\ket{p_0,p_1,q_0,q_1;\overline{p}_0,\overline{q}_0}$ basis and $(p_1p_0)_2 \times (q_1q_0)_2$ basis, respectively.}
\label{Fig1}
\end{figure*}

We describe the method of programming the integer factorization problem with a Rydberg-atom graph, using the simplest possible example of $p\times q=6$ as illustrated in Fig.~\ref{Fig1}.
Initially, we reduce the integer factorization problem to a SAT one. Considering the three-bit binary representation $6=(110)_2$, we assume that the factors $p$ and $q$ are two-bit integers, denoted as $p=(p_1 p_0)_2$ and $q=(q_1 q_0)_2$. The Boolean equations governing the binary variables $p_0, p_1, q_0, q_1$ are then derived from the multiplication table in Fig.~\ref{Fig1}(a) as follows:
\begin{subequations} \label{Factor6}
\begin{eqnarray}
p_0 q_0 &=&0, \label{Factor6Begin_0}\\
p_0 q_1\oplus p_1 q_0 &=& 1, \label{Factor6Begin_1}\\
p_0 q_1 p_1 q_0\oplus p_1 q_1 &=& 1, \label{Factor6Begin_2}
\end{eqnarray} 
\end{subequations}
where $\oplus$ is XOR and $p_0q_1p_1q_0$ in Eq.~\eqref{Factor6Begin_2} denotes the carry arising from Eq.~\eqref{Factor6Begin_1}. These equations can be efficiently (i.e., in a polynomial number of steps in the bit number of $n$) converted to a Boolean equation in conjunctive normal form, yielding
\begin{equation}
\Psi_6= p_1q_1(p_0+q_0)(\bar{p}_0+\bar{q}_0) =1.
\label{CNF6}
\end{equation}
Further details will be described in Sec.~\ref{SecIII}. The Boolean equation $\Psi_6=1$ readily translates into the following Boolean satisfiability (SAT) problem of 4 clauses:
\begin{subequations} \label{Psi6}
\begin{eqnarray}
\Psi_6(p_0,p_1,q_0,q_1) &=& C_1 \wedge C_2 \wedge C_3 \wedge C_4=1, \\
C_1 &=& p_1, \label{Psi61} \\
C_2 &=& q_1, \label{Psi62} \\
C_3 &=& p_0 \vee q_0 , \label{Psi63} \\
C_4 &=& \overline{p}_0 \vee \overline{q}_0. \label{Psi64}
\end{eqnarray}
\end{subequations}

Subsequently, we translate this SAT problem into an MIS problem on a graph, as depicted in Fig.~\ref{Fig1}(c). The first two clauses $C_1$ and $C_2$ in Eqs.~\eqref{Psi61} and \eqref{Psi62} are incorporated into isolated single-vertex graphs denoted as $\mathbf{K}_1$'s in graph nomenclature:
\begin{equation}
\mathbf{K}_1 (v) := G(V=\{v\},E=\emptyset).
\end{equation}
For $C_3$ and $C_4$, two-vertex connected graphs, or $\mathbf{K}_2$'s, are employed:
\begin{equation}
\mathbf{K}_2 (v_1,v_2) := G(V=\{v_1, v_2\}, E= \{(v_1,v_2)\}.
\end{equation}
Now, we introduce additional edges to represent inter-clause relations between variables $p_0$, $q_0$ and their negations $\overline{p}_0$, $\overline{q}_0$, creating connections between $C_3$ and $C_4$. These inter-clause edges impose constraints between the vertices encoding the same variables. Consequently, the graph ${G}_6$ expressing the factorization problem $p\times q=6$ as an MIS problem on the graph is defined as follows:
\begin{eqnarray} \label{G6}
G_{6}&=&\mathbf{K}_1 (p_1) \cup \mathbf{K}_1 (q_1) \cup \mathbf{K}_2 (p_0,q_0) \cup \mathbf{K}_2 (\overline{p}_0,\overline{q}_0) \nonumber \\
&+&\mathbf{e}(p_0,\overline{p}_0)+\mathbf{e}(q_0,\overline{q}_0). \label{G6}
\end{eqnarray}
Here, $G_6$ has maximum independent set of size 4, which is equal to the number of clauses in $\Psi_6$. Therefore, any maximum independent set configuration of size 4 corresponds to an assignment satisfying $\Psi_6=1$, ensuing that the corresponding binary representation of numbers $p,q$ meet $p\times q = 6$.

By transforming the integer factorization problem into an MIS one, we can leverage a Rydberg-atom experiment to address it, exploiting the Rydberg blockade phenomenon to inherently encode the independence condition in the spatial configuration of the atoms. Specifically, the graph $G_6$ is implemented as a Rydberg-atom graph, where each vertex corresponds to an atom, and edges are established by positioning the respective pairs of atoms in close proximity, ensuring that their simultaneous excitation to the Rydberg state is hindered by the blockade phenomenon~\cite{Rydberg_blockade_urban,Rydberg_blockade_geatan}. Following a quantum evolution, the collection of atoms in the Rydberg state delineates a subset of vertices of $G_6$. These subsets sampled from this final state are expected to be good candidates for the MIS of the graph.

The Hamiltonian is defined for a general graph $G$ (an unweighted graph) as follows:
\begin{equation}
\hat H(G)= \frac{U}{4}\sum_{(j,k)\in E} (\hat \sigma_z^{(j)}+1)(\hat\sigma_z^{(k)}+1)-\frac{\hbar \Delta }{2} \sum_{j\in V} \hat\sigma_z^{(j)},
\end{equation}
where $\ket{0}$ and $\ket{1}$ are pseudo-spin states denoting the ground and Rydberg-atom states, respectively, $\hat \sigma_z=\ket{1}\bra{1}-\ket{0}\bra{0}$, $U$ is the interaction between each pair of ``edged'' atoms (of the same separation distance), and $\Delta$ is the detuning of Rydberg-atom excitation. The MIS phase requires two conditions: $U\gg \hbar|\Delta|$ enforces the Rydberg blockade phenomenon; and $\Delta>0$ maximizes the number of atoms being excited to the Rydberg state. The many-body ground state, $\ket{\hat H(G)}$, is then the superposition of MIS's of $G$~\cite{Pichler2018,Minhyuk2022}. For the Rydberg-atom graph $G_6$ in Eq.~\eqref{G6}, the many-body ground state $\ket{\hat H({G_6})}$ of $\hat H({G_6})$ is given by:
\begin{equation} \label{HG6}
\ket{\hat H({G_6})(\ket{p_0p_1q_0q_1;\overline{p}_0\overline{q}_0})}
=\frac{\ket{0111;10}+\ket{1101;01}}{\sqrt{2}}, 
\end{equation}
which is the superposition of two factoring solutions, $p \times q=(10)_2 \times (11)_2$ and $(11)_2 \times (10)_2$. 

Experimental verification can be performed with the adiabatic evolution of the Rydberg-atom graph $G_6$ from the paramagnetic phase to the anti-ferromagnetic phase, which corresponds to the MIS phase~\cite{Jeong2023}. We used rubidium atoms ($^{87}\mathrm{Rb}$) with ground state $\ket{0}\equiv \ket{5S_{1/2},F=2, m_F=2}$ and Rydberg state  $\ket{1}\equiv \ket{71S_{1/2},m_J=1/2}$. The Rabi frequency $\Omega$ is ramped up from 0 to $\Omega_0=(2\pi)1.5$~MHz, 
while the laser detuning is maintained at $\Delta=-(2\pi)3.5$~MHz for 0.3~$\mu$s. Then the detuning is ramped up from $-(2\pi)3.5$ to $+(2\pi)\delta_F$~MHz for 2.4~$\mu$s, with $\delta_F=3.5$ and fixed Rabi frequency $\Omega_0$. (The $p\times q=15$ and $35$ experiments in Sec.~\ref{SecIV} are performed with $\delta_F=3.5$ and 3.9, respectively.) Finally the Rabi frequency is ramped down to zero and the detuning is maintained at $+(2\pi)3.5$~MHz for 0.3~$\mu$s. The entire evolution time is 3.0~$\mu$s. The detuning and Rabi frequencies are changed with the frequency and the power of the excitation lasers with acousto-optic modulators (AOM), which are controlled by a programmable radio-frequency synthesizer (Moglabs XRF). After the quasi-adiabatic evolution, the population of each atom is measured by illuminating the conventional cycling transition lights where the atoms in the ground state show fluorescence whereas the atoms in the Rydberg state do not. The experimental results obtained with $G_6$ are depicted in Fig.~\ref{Fig1}(d), where the expected states $\ket{p_0p_1q_0q_1;\overline{p}_0\overline{q}_0}=\ket{0111;10}$ and $\ket{1101;01}$ in Eq.~\eqref{HG6} are measured with high probabilities, confirming that the integer factors are $(11)_2=3$ and $(10)_2=2$.

\section{Encoding integer factorization into the satisfiability problem}\label{SecIII}
The conjunctive normal form of integer multiplication can be efficiently represented using a binary decision diagram (BDD), as detailed in Ref.~\cite{Raddum2019}. In this context, we provide a concise overview of the BDD construction process, utilizing the example of $p\times q=6$ in Sec.~\ref{SecII}, and derive $\Psi_6$ in Eq.~\eqref{Psi6}, the Boolean expression in conjunctive normal form for the SAT problem~\cite{Russel1995,Srebrny2004}.

In a generic factorization problem for a given semiprime number $n = p \times q$, if $n$ is an $N$-bit integer, finding the unknown factors, $p$ (an $N_p$-bit integer) and $q$ (an $N_q$-bit integer), involves identifying the preimage of $n$ through the binary multiplication function 
\begin{equation}
f: \{0,1\}^{N_p} \times \{0,1\}^{N_q} \rightarrow \{0,1\}^N,
\end{equation}
where $N_q \ge N-N_p$. Let $p_i \,(0\le i \le N_p-1)$ be the $i$-th bit of $p$, $q_j \, (0\le j \le N_q-1)$ the $j$-th bit of $q$, and $n_k \, (0\le k \le N -1)$ the $k$-th bit of $n$. The function $f$ is then expressed as:
\begin{eqnarray} \label{eq:factorisation_as_sum}
 n= f(p_0,p_1,\cdots, q_0, q_1, \cdots)  = \sum_{i=0}^{N_p -1} \sum_{j=0}^{N_q-1} p_i q_j 2^{i+j},
\end{eqnarray}
where each term $p_i q_j 2^{i+j}$ contributes a nonzero value to the sum only when $p_i=q_j=1$. 

\begin{figure}[ht!]
    \centering
\includegraphics[width=0.45\textwidth]{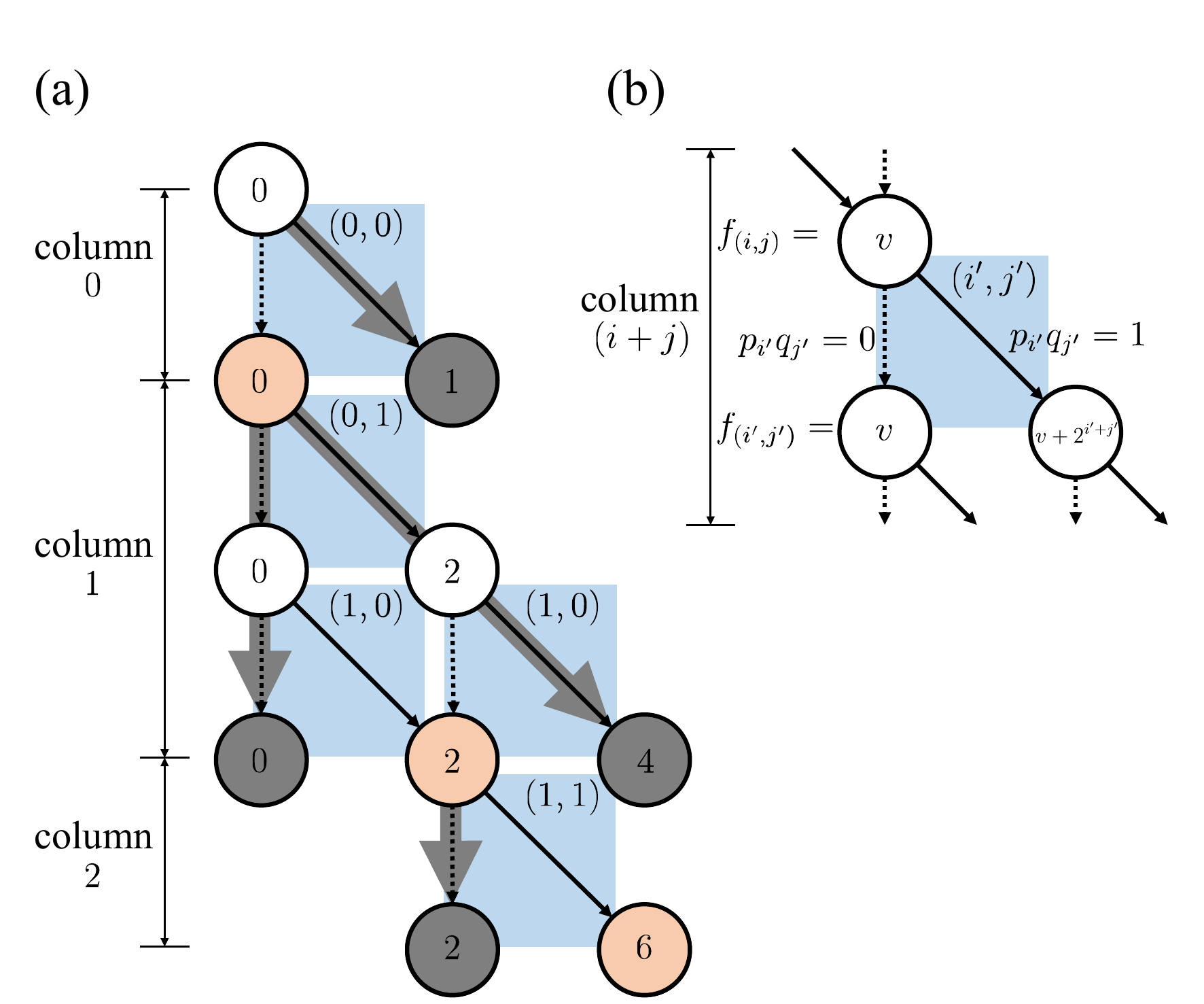}
    \caption{(a) The binary decision diagram (BDD) for factoring $p\times q=6$. (b) The unit BDD representing the logical relation between $p_{i'}$ and $q_{j'}$ in the multiplication table, depicted by the blue-colored box labeled  $f_{(i',j')}$, where the edges in the BDD can be either solid or dotted, depending on whether $p_{i'}q_{j'}$ is 1 or 0.}
\label{Fig2}
\end{figure}

Figure~\ref{Fig2}(a) shows the BDD for the $p\times q=6$ factoring problem. In this constructed BDD, each column corresponds to one of the three bit-wise calculations of $p\times q=6$ in Eq.~\eqref{Factor6}. There are three columns comprising a total of 10 unit BDDs, interconnected based on the constraints from bit-wise equations in Eq.~\eqref{eq:factorisation_as_sum}. Assuming that we have processed the sum in Eq.~\eqref{eq:factorisation_as_sum} up to the $(i,j)$-th term, by defining a running sum $f_{(i,j)}=v$, the subsequent running sum $f_{(i',j')}$ is given by:
\begin{equation} \label{eq:ij-term}
f_{(i',j')}(p_{i'}, q_{j'}; v) = \begin{cases}
	v &\text{if } p_{i'}q_{j'} = 0, \\
    	v + 2^{i'+j'} & \text{if } p_{i'} =q_{j'} = 1.
\end{cases}
\end{equation}
These running sum relations can be represented by the unit BDD, as depicted in Fig.~\ref{Fig2}(b). The starting and ending nodes contains the running values $f_{(i,j)}$ and $f_{(i',j')}$, respectively, with edges distinguished as solid or dotted based on whether the product $p_{i'}q_{j'}$  is set to 1 or 0, respectively. The BDD for $f$ is formed by linking these unit BDDs representing all running sum relations. 

The first column (column 0) in Fig.~\ref{Fig2}(a) pertains to the first bit calculation, utilizing the topmost unit BDD to compute the running sum $f_{(0,0)}(p_0,q_0)=p_0q_0$. Two possible end node values, $f_{(0,0)}=0$ (derived from paths satisfying $p_0q_0=0$) and  $f_{(0,0)}=1$ (through the path of $p_0q_0=1$, ), emerge. Only the former (the orange-colored, left end node) aligns with the first bit constraint $f_{(0,0)}=n_0$, leading to the exclusion of the latter (the gray-colored, right end node). The second column (column 1) initiates from the $f_{(0,0)}=0$ node and computes the second bit of $n$. Three possible end nodes represent the running sums $f_{(1,0)}=0$, 2, and 4, respectively. Among these, only the second one satisfies the second bit equation, $f_{(1,0)}=n_0+n_1\times 2$.  Similarly, the third column (column 2) begins from the $f_{(1,0)}=2$ node and terminates at two possible nodes, $f_{(1,1)}=2$ and $6$. The second one is the only one satisfying $f_{(1,1)}=n_0+n_1\times 2+n_2 \times 2^2$. 

Now we seek a Boolean expression in conjunctive normal form for the 3-SAT problem. Generally, the prime number couple $(p,q)$ for a semi-prime number $n=p\times q$ is unique (up to ordering). So, the determination of BDD paths leading to the end node corresponding to $n$ establishes the values of $p_i$s and $q_j$s. Since $p_i$s and $q_j$s are involved in various paths, fixing their values imposes constraints on the BDD paths. Conversely, preventing certain paths from extending beyond the solution space by incorrectly setting a bit-wise equation for $n_i$ introduces constraints on the potential values of $p_i$s and $q_j$s. Aggregating these constraints enables the representation of the given  factorization problem through a Boolean formula of the SAT problem.

In the illustrated BDD example in Fig.~\ref{Fig2}(a), the paths (gray arrows) unable to meet the constraints are as follows: $p_0 q_0=1$ in column 0, $\overline{p_0 q_1} \cdot \overline{p_1 q_0} = 1$ and $p_0 q_1 p_1 q_0 =1 $ in column 1, and $q_2 =1$, $\overline{p_1 q_1}=1$, and $p_2=1$ in column 2. These unsuccessful paths, which are the gray-colored arrows in Fig.~\ref{Fig2}(b), can be expressed in disjunctive norm form as follows:
\begin{subequations}
\begin{eqnarray}
\neg \Psi_{6,0}&=& p_0 q_0, \\
\neg \Psi_{6,1}&=&\overline{p_0 q_1} \cdot \overline{p_1 q_0}+p_0 q_1 p_1 q_0  \\ 
\neg \Psi_{6,2}&=& \overline{p_1q_1}
\end{eqnarray}
\end{subequations}
The resulting Boolean equation for $p\times q=6$ is given by 
\begin{eqnarray}
\Psi_{6}&=&\neg(\neg \Psi_{6,0}+\neg \Psi_{6,1} +\neg\Psi_{6,2}) \nonumber \\
&=& p_1 q_1 \left( p_0 + q_0 \right) \left( \overline{p}_0 + \overline{q}_0 \right)  =1,
\end{eqnarray}
which is the same as $\Psi_6$ in Eq.~\eqref{CNF6} in Sec.~\ref{SecII}.

\section{Experimental demonstration} \label{SecIV} 
We now apply the process of translating factoring problems into Rydberg-atom graphs to experimentally investigate the integer factors of $p\times q=15$ and $p\times q=35$. We first derive the corresponding Boolean expressions in conjunctive normal form using the methodology outlined in Sec.~\ref{SecIII} and subsequently these expressions are transformed into their respective Rydberg-atom graphs. 

\subsection{Solving $p\times q = 15$}

\begin{figure*}[thbp]
  \centering
\includegraphics[width=0.98\textwidth]{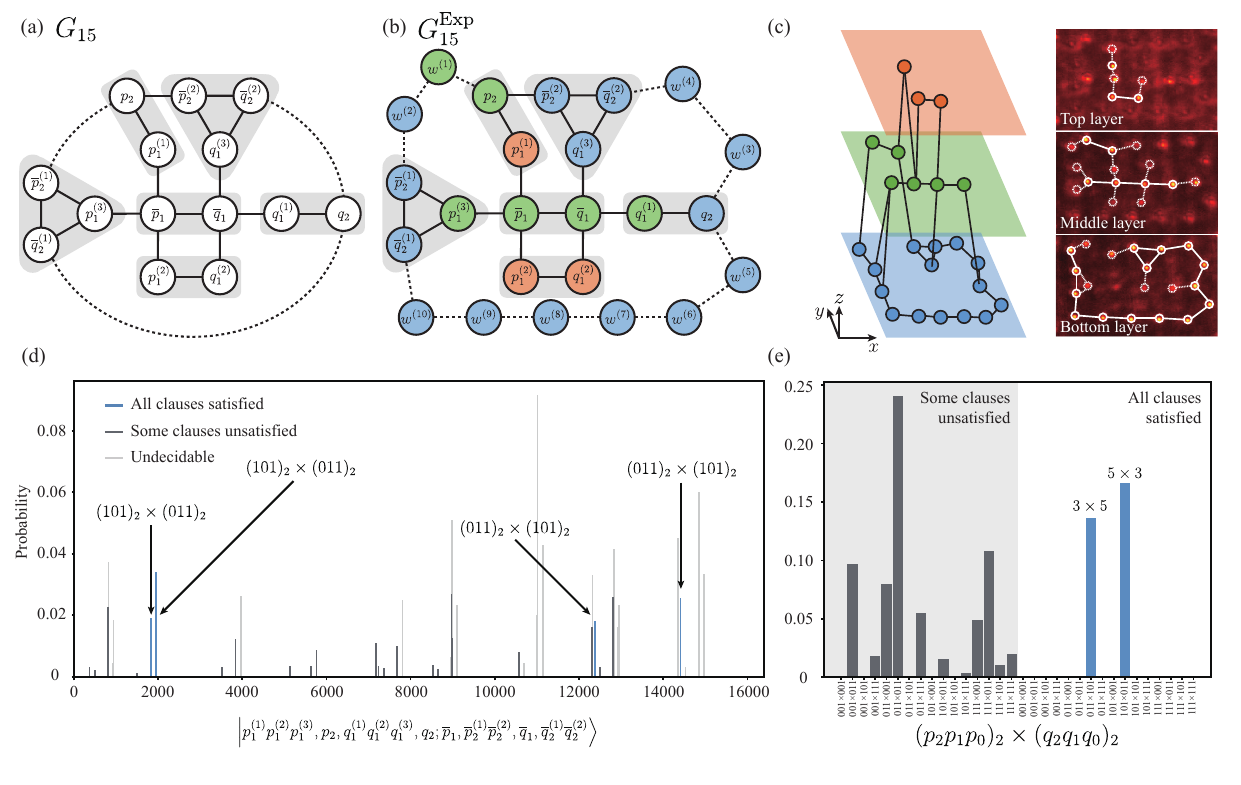}
\caption{Implementing the integer factorization of $p\times q=15$ using Rydberg atoms. (a) Rydberg-atom graph $G_{15}$. (b) Experimental Rydberg-atom graph $G_{15}^{\mathrm{Exp}}$ and (c) image of atoms in 3D configuration. (d) Experimental probability distribution for all microstates in the basis of variable and negation atoms of $G^{\rm Exp}_{15}$. The four blue peaks are $\ket{000,1,111,0;100010}$ corresponding to $((p_2 p_1 p_0)_2,(q_2 q_1 q_0)_2)=(101;011)$, $\ket{111,0,000,1;0,01,1,00}$ corresponding to $(011;101)$, $\ket{000,1,110,0;1,00,0,11}$ corresponding to $(101;011)$, and $\ket{110,0,000,1;0,11,1,00}$  corresponding to $(011;101)$. (e) Result of factorization. The portion of peaks in (d) following the 3-SAT logical constraints is mapped to $(011;101)=(3,5)$ and $(101,011)=(5,3)$, demonstrating the correct answer to the factoring problem $15=3\times 5$.
}
\label{Fig3}
\end{figure*}

The conjunctive normal form Boolean equation representing $p \times q=15$ is derived as follows:
\begin{eqnarray} \label{Psi15a}
\Psi_{15}
&=&(p_1+p_2)(q_1+q_2)(p_1+q_1)(\overline{p}_1+\overline{q}_1)\nonumber \\
   &&\cdot (p_1+\overline{p}_2+\overline{q}_2)(q_1+\overline{p}_2+\overline{q}_2)p_0q_0=1.
\end{eqnarray}
The detailed construction of $\Psi_{15}$ from the BDD is described in Appendix. The corresponding Rydberg-atom graph $G_{15}$ is then expressed as:
\begin{eqnarray} \label{G15}
G_{15} &=& \mathbf{K}_2(p_1^{(1)},p_2) \cup \mathbf{K}_2 (q_1^{(1)},q_2) \cup \mathbf{K}_2 (p_1^{(2)},q_1^{(2)})  \nonumber \\
 &\cup  & \mathbf{K}_2(\overline{p}_1,\overline{q}_1) \cup \mathbf{K}_3 (p_1^{(3)},\overline{p}_2^{(1)}, \overline{q}_2^{(1)}) \nonumber \\
 & \cup & \mathbf{K}_3 (q_1^{(3)}, \overline{p}_2^{(2)},\overline{q}_2^{(2)})  \nonumber \\
 &+& \sum_{i=1,2,3}  \mathbf{e}(p_1^{(i)}, \overline{p}_1)+ \sum_{j=1,2}   \mathbf{e}(p_2,\overline{p}_2^{(j)}) \nonumber \\
 &+& \sum_{i=1,2,3} \mathbf{e}(q_1^{(i)},\overline{q}_1) + \sum_{j=1,2} \mathbf{e}(q_2,\overline{q}_2^{(j)}), 
\end{eqnarray}
which is depicted in Fig.~\ref{Fig3}(a). Each parenthesized term in $\Psi_{15}$ corresponds to a two- or three-vertex connected graph ($\mathbf{K}_2$ or $\mathbf{K}_3$), and each variable-negation relation is represented by an additional edge. The superscript indices in the last four terms in Eq.~\eqref{G15} denote variable-atom duplicates in different subgraphs ($\mathbf{K}_2$ or $\mathbf{K}_3$). The single-vertex graphs for $p_0$ and $q_0$ are omitted in Fig.~\ref{Fig3}(a), for simplicity. 

\begin{table*}[t] 
\caption{Atom positions of $G_{15}^{\mathrm{Exp}}$, $G_{35}^{\mathrm{Exp}}$}
\begin{ruledtabular}
\begin{tabular}{@{}cllllllll@{}}
Graphs & \multicolumn{4}{c}{Atom positions $(x,y,z)$ or $(x,y)$ ($\mu$m)}   \\
\hline
\multirow{4}{*}{$G_{15}^{\mathrm{Exp}}$} 
& $p_1^{(1)}$:& $(0.0,6.43,6.43)$ & $p_1^{(2)}$:& $(0.0,-6.43,6.43)$ & $p_1^{(3)}$: & $(-9.09,0.0,0.0)$ \\
&  $p_2$: & $(-1.64,12.67,0.0)$ & $q_1^{(1)}$: & $(18.18,0.0,0.0)$ &  $q_1^{(2)}$: & $(9.09,-6.43,6.43)$ \\
& $q_1^{(3)}$: & $(9.09,6.43,-6.43)$ & $q_2$:& $(24.61,0.0,-6.43)$ & $\overline{p}_1$: & $(0.0,0.0,0.0)$ \\
& $\overline{p}_2^{(1)}$:& $(-13.64,4.55,-6.43)$ & $\overline{p}_2^{(2)}$:& $(4.55,14.3,-6.43)$ & $\overline{q}_1$:& $(9.09,0.0,0.0)$ \\
& $\overline{q}_2^{(1)}$: & $(-13.64,-4.55,-6.43)$  & $\overline{q}_2^{(2)}$:& $(13.64,14.3,-6.43)$ & $w^{(1)}$: & $(-10.0,16.31,0.0)$\\& $w^{(2)}$:& $(-15.45,12.73,-6.43)$ & $w^{(3)}$: & $(27.34,8.67,-6.43)$ & $w^{(4)}$:& $(22.18,15.85,-6.43)$ \\
& $w^{(5)}$: & $(29.61,-7.0,-6.43)$  & $w^{(6)}$:& $(22.73,-13.64,-6.43)$ & $w^{(7)}$: & $(13.64,-13.64,-6.43)$ \\
& $w^{(8)}$:& $(4.55,-13.64,-6.43)$ & $w^{(9)}$: & $(-4.55,-13.64,-6.43)$ & $w^{(10)}$:& $(-13.64,-13.64,-6.43)$\\
\hline
\multirow{5}{*}{$G_{35}^{\mathrm{Exp}}$}  
& $p_1^{(1)}$:& (35.36,32.14) & $p_1^{(2)}$:& (36.43,43.21) & $p_1^{(3)}$: &(78.21,41.43) \\
& $p_2^{(1)}$: &(29.29,27.14)  & $p_2^{(2)}$:& (62.14,29.29) & $q_1^{(1)}$:  &(62.14,48.21) \\
& $q_1^{(2)}$:  &(60.36,36.79) & $q_2^{(1)}$: &(42.50,48.21) & $q_2^{(2)}$: &(53.93,47.14) \\
& $\overline{p}_1^{(1)}$: &(42.14,37.5) & $\overline{p}_1^{(2)}$: &(56.43,53.93) & $\overline{p}_1^{(3)}$:  &(73.21,47.14) \\
& $\overline{q}_1$:  &(66.79,41.79) & $\overline{q}_2$: &(47.86,42.14)  & $q_1^{(3)}$: &(71.79,36.43) \\
& $s$: & (55.36,33.93) & $\overline{s}$:& (48.93,34.29)   & $w^{(1)}$:  &(84.29,47.86) \\
 & $w^{(2)}$: &(80.00,55.00) & $w^{(3)}$:  &(73.57,61.07)  & $w^{(4)}$: &(64.29,59.29)\\
 & $w^{(5)}$:  &(31.43,50.36) & $w^{(6)}$: &(34.64,57.50) & $w^{(7)}$:  &(42.50,61.43) \\
 & $w^{(8)}$: &(50.36,61.07) &  \\
\end{tabular}
\end{ruledtabular}
\label{Table1}
\end{table*}

The graph $G_{15}$ is not directly implementable with a two-dimensional (2D) arrangement of atoms. To address this, we opt for a three-dimensional (3D) atomic arrangement, following the approach outlined in Ref.~\cite{Minhyuk2020,YHSong2021}. The resulting experimental Rydberg-atom graph $G^{\rm Exp}_{15}$ is illustrated in Fig.~\ref{Fig3}(b). The graph features a three-layer atomic structure: the top layer contains three variables, $p_1^{(1)}, p_1^{(2)}, q_1^{(2)}$; the middle layer encompasses five variables, $p_2$, $\overline{p}_1$, $p_1^{(3)}, \overline{q}_1, q_1^{(1)}$; and the bottom layer includes the remaining six variables, $\overline{p}_2^{(1)}, \overline{q}_2^{(1)}, q_1^{(3)}, \overline{p}_2^{(2)}, \overline{q}_2^{(2)}, q_2$. Three Rydberg quantum wires operate on the middle and bottom layers: $\mathbf{P}_4(p_2,w^{(1)},w^{(2)},\overline{p}_2^{(1)})$ for $\mathbf{e}(p_2,\overline{p}_2^{(1)})$, $\mathbf{P}_4(q_2,w^{(3)},w^{(4)},\overline{q}_2^{(2)})$ for $\mathbf{e}(q_2,\overline{q}_2^{(2)})$, and $\mathbf{P}_8(q_2,w^{(5)},w^{(6)},\cdots,w^{(10)},\overline{q}_2^{(1)})$ for $\mathbf{e}(q_2,\overline{q}_2^{(1)})$. The total number of atoms in $G^{\rm Exp}_{15}$ is 24, comprising 14 variable atoms ($p_1^{(1)}$, $p_2$, $q_1^{(1)}$, etc.) and 10 quantum-wire atoms ($w^{(1)}$, $w^{(2)}$, etc.). The atom images of $G^{\rm Exp}_{15}$ are presented in Fig.~\ref{Fig3}(b) and the 3D coordinates of all atoms are detailed in Table~\ref{Table1}.

Experimental results of $G^{\rm Exp}_{15}$ are shown in Fig.~\ref{Fig3}(c). The quantum adiabatic evolution of the Hamiltonian $H(G_{15}^{\rm Exp})$ is performed, following the experimental procedure outlined in Sec.~\ref{SecII}. A total of $12,575$ experimental events were recorded. We applied the single-vertex error mitigation protocol from Ref.~\cite{EbadiScience2022_postprocessing} and the Rydberg quantum-wire compilation method of anti-ferromangetic
chain states from Ref.~\cite{Minhyuk2022} to obtain 11, 274 usable events. The resulting probability distribution is depicted in Fig.~\ref{Fig3}(d) for all  microstates in the basis $\ket{ p_1^{(1)}p_1^{(2)}p_1^{(3)}, p_2, q_1^{(1)} q_1^{(2)} q_1^{(3)}, q_2; \overline{p}_1, \overline{p}_2^{(1)} \overline{p}_2^{(2)},  \overline{q}_1,  \overline{q}_2^{(1)} \overline{q}_2^{(2)}}$ covering all variable and negation atoms of $G^{\rm Exp}_{15}$. In Fig.~\ref{Fig3}(d), the experimental probability distribution $P(p,q)$ is presented in the $((p_2 p_1 p_0)_2;(q_2 q_1 q_0)_2)$ basis. The standard variable assignment method for duplicate variables~\cite{Vazirani} was used, wherein a variable $x$ is binary one if any duplicate $x^{(i)}$ is one, and all of its negations $\overline{x}^{(j)}$s are zero at the same time. Conversely, it is binary zero if all duplicates $x^{(i)}$ are zero, and any variable duplicate $\overline{x}^{(j)}$ is one. If neither condition is met, it is marked as ``undecidable.'' As a result, the Rydberg-atom experiment of $G^{\rm Exp}_{15}$ determines the integer factors of $p \times q=15$ to be $(p,q)=(3,5)$ and $(5,3)$ with 13.4\% and 16.6\% probabilities, respectively.

\subsection{Solving $p\times q = 35$}
\begin{figure*}[thbp]
  \centering
\includegraphics[width=0.98\textwidth]{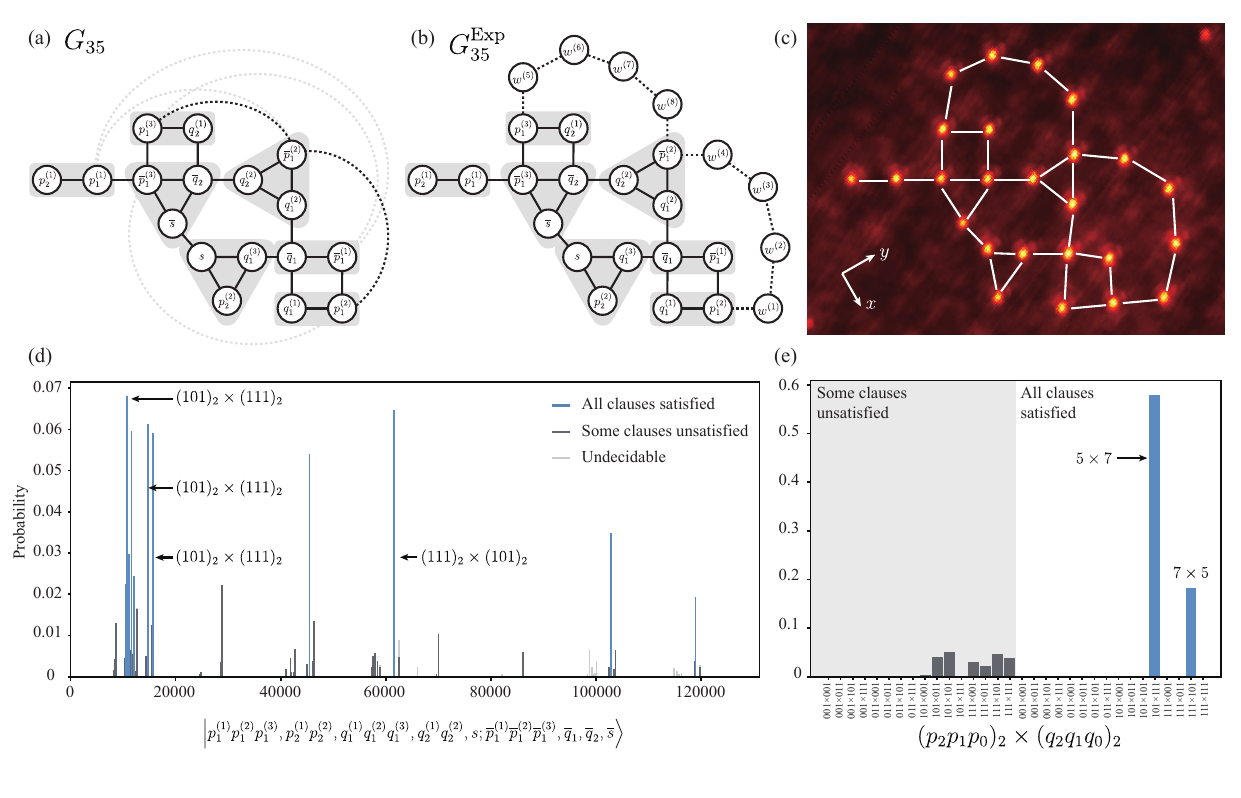}
\caption{Implementing the integer factorization of $p\times q=35$ using Rydberg atoms. (a) Rydberg-atom graph $G_{35}$. (b) Experimental Rydberg-atom graph $G_{35}^{\mathrm{Exp}}$ and (c) the image of the atom configuration. (d) Probability distribution obtained experimentally for all microstates in the basis of variable and negation atoms in $G^{\rm Exp}_{35}$. The top four most probable microstates in (d) are $\ket{000,10,100,11,1;101,0,0,0}$ 
corresponding to $((p_2 p_1 p_0)_2, (q_2 q_1 q_0)_2)=(101, 111)$; $\ket{000,11,100,11,0;100,0,0,1}$ 
corresponding to $(101, 111)$; $\ket{011,11,000,01,0;000,1,0,1}$ 
corresponding to $(111, 101)$; and $\ket{000,11,110,10,0;100,0,0,1}$ 
corresponding to $(101, 111)$. (e) Result of factorization. The portion of peaks in (d) following the 3-SAT logical constraints is mapped to $(101;111)=(5,7)$ and $(111;101)=(7,5)$, demonstrating the correct answer to the factoring problem $35=5\times 7$. }
\label{Fig4}
\end{figure*}

Let us examine another example: $p\times q=35$. The procedure is similar to the previous $p\times q=15$ example. The SAT formula $\Psi_{35}$ is derived as follows:
\begin{eqnarray}
\Psi_{35}&=&p_0q_0(p_1+p_2)(p_1+q_1) (p_1+q_2)(\overline{p}_1+\overline{q}_1) \nonumber \\
&&\cdot (q_1+q_2+\overline{p}_1) (p_2+q_1+\overline{p}_1+\overline{q}_2)=1. \label{Psi35a}
\end{eqnarray}
It can be reformulated for 3-SAT implementation as:
\begin{eqnarray}
\Psi_{35}&=&p_0q_0(p_1+p_2)(p_1+q_1) (p_1+q_2)(\overline{p}_1+\overline{q}_1) \nonumber \\
&&\cdot (q_1+q_2+\overline{p}_1) (p_2+q_1+s) \nonumber \\
&&\cdot (\overline{p}_1+\overline{q}_2+\overline{s}) =1. \label{Psi35}
\end{eqnarray}
This involves replacing the last clause $(p_2+q_1+\overline{p}_1+\overline{q}_2)$ in Eq.~\eqref{Psi35a}, which has four variables, with $(p_2+q_1+s)(\overline{p}_1+\overline{q}_2+\overline{s})$ by introducing a dummy variable $s$. The corresponding Rydberg-atom graph $G_{35}$ is given by
\begin{eqnarray} \label{G35}
G_{35}&=& \mathbf{K}_2(p_1^{(1)},p_2^{(1)}) \cup \mathbf{K}_2(p_1^{(2)}, q_1^{(1)}) \cup \mathbf{K}_2(p_1^{(3)},q_2^{(1)})  \nonumber \\
 &\cup& \mathbf{K}_2(\overline{p}_1^{(1)},\overline{q}_1) \cup \mathbf{K}_3(q_1^{(2)},q_2^{(2)},\overline{p}_1^{(2)})  \nonumber \\
 &\cup&   \mathbf{K}_3 (p_2^{(2)},q_1^{(3)},s) \cup \mathbf{K}_3(\overline{p}_1^{(3)},\overline{q}_2,\overline{s}) \nonumber \\
&+& \sum_{i,j=1}^3 \mathbf{e}(p_1^{(i)},\overline{p}_1^{(j)}) 
+\sum_{i=1}^3\mathbf{e}(q_1^{(i)},\overline{q}_1) +\sum_{i=1}^2\mathbf{e}(q_2^{(i)},\overline{q}_2) \nonumber \\
&+&\mathbf{e}(s,\overline{s}),
\end{eqnarray}
as depicted in Fig.~\ref{Fig4}(a). 

Implementing $G_{35}$ directly in an experiment is challenging, even with Rydberg quantum wires. Hence, we explore a 2D version of the experimental graph, denoted as $G_{35}^{\rm Exp}$, expressed by the following:
\begin{eqnarray}
G_{35}^{\rm Exp}&=& 
\mathbf{K}_2(p_1^{(1)},p_2^{(1)}) \cup \mathbf{K}_2(p_1^{(2)}, q_1^{(1)}) \cup \mathbf{K}_2(p_1^{(3)},q_2^{(1)})  \nonumber \\
 &\cup& \mathbf{K}_2(\overline{p}_1^{(1)},\overline{q}_1) \cup \mathbf{K}_3(q_1^{(2)},q_2^{(2)},\overline{p}_1^{(2)})  \nonumber \\
 &\cup&   \mathbf{K}_3 (p_2^{(2)},q_1^{(3)},s) \cup \mathbf{K}_3(\overline{p}_1^{(3)},\overline{q}_2,\overline{s}) \nonumber \\
  &\cup&\mathbf{P}_6(p_1^{(2)},w^{(1)},w^{(2)},w^{(3)},w^{(4)},\overline{p}_1^{(2)})\nonumber\\
 &\cup&\mathbf{P}_6(p_1^{(3)},w^{(5)},w^{(6)},w^{(7)},w^{(8)},\overline{p}_1^{(2)}) \nonumber \\
&+& \mathbf{e}(p_1^{(1)},\overline{p}_1^{(3)})+\mathbf{e}(p_1^{(2)},\overline{p}_1^{(1)})+\mathbf{e}(p_1^{(3)},\overline{p}_1^{(3)}) \nonumber\\
&+& \sum_{i=1}^3\mathbf{e}(q_1^{(i)},\overline{q}_1) +\sum_{i=1}^2\mathbf{e}(q_2^{(i)},\overline{q}_2) \nonumber \\
&+&\mathbf{e}(s,\overline{s}),
\end{eqnarray}
as illustrated in Fig.~\ref{Fig4}(b). In this representation, the term $\sum_{i,j=1}^3 \mathbf{e}(p_1^{(i)},\overline{p}_1^{(j)})$ in Eq.~\eqref{G35} is replaced by $\mathbf{e}(p_1^{(1)},\overline{p}_1^{(3)})+\mathbf{e}(p_1^{(2)},\overline{p}_1^{(1)})+\mathbf{e}(p_1^{(3)},\overline{p}_1^{(3)})$. Two Rydberg quantum wires $\mathbf{P}_6(p_1^{(2)},\cdots,\overline{p}_1^{(2)})$ and $\mathbf{P}_6(p_1^{(3)},\cdots,\overline{p}_1^{(2)})$ are introduced for edges $\mathbf{e}(p_1^{(2)},\overline{p}_1^{(2)})$ and $\mathbf{e}(p_1^{(3)}, \overline{p}_1^{(2)})$, respectively. The remaining four edges, $\mathbf{e}(p_1^{(1)}, \overline{p}_1^{(1)})$, $\mathbf{e}(p_1^{(1)}, \overline{p}_1^{(2)})$, $\mathbf{e}(p_1^{(2)}, \overline{p}_1^{(3)})$, $\mathbf{e}(p_1^{(3)}, \overline{p}_1^{(1)})$, are treated post-selectively. This 2D graph involves a total of 25 atoms, comprising 17 variable atoms and 8 quantum-wire atoms. The atom image is presented in Fig.~\ref{Fig4}(b) and the the 2D coordinates of all atoms are detailed in Table~\ref{Table1}. 

Figure~\ref{Fig4}(d) showcases the experimental outcomes of $G^{\rm Exp}_{35}$. The probability distribution is plotted in the basis 
of all variable and negation atoms of $G_{35}^{\mathrm{Exp}}$. Among the 13 microstates highlighted in blue, there are correct mapping to the solution factors.  Conversely, the microstates labeled in black are deemed $\Psi_{35}$-unsatisfiable, and those in gray are considered ``undecidable'' regarding a definite set of values for variables $(p_2, p_1, q_2, q_1, s)$. Out of 9,383 experimental events, 5,337 usable events were collected after the procedures of the single-vertex error mitigation protocol~\cite{EbadiScience2022_postprocessing} and the Rydberg quantum-wire compilation method~\cite{Minhyuk2022} are applied. In contrast to the previous $p \times q = 15$ experiment, where 89.7\% of total events are usable, the case of $p\times q=35$ yields only 56.9\% usable events, reflecting the consequence of post-selective treatment of the four edges $\mathbf{e}(p_1^{(1)}, \overline{p}_1^{(1)})$, $\mathbf{e}(p_1^{(1)}, \overline{p}_1^{(2)})$, $\mathbf{e}(p_1^{(2)}, \overline{p}_1^{(3)})$, $\mathbf{e}(p_1^{(3)}, \overline{p}_1^{(1)})$ in $G_{35}$, but not in $G_{35}^{\mathrm{Exp}}$. In Fig.\ref{Fig4}(e), the probability distribution for solution factors $(p,q)$ is visually presented, employing a color-coding scheme. Probability peaks corresponding to the correct solution factors are highlighted in blue, those for microstates failing to satisfy $\Psi_{35}$ are depicted in black, and the probabilities for "undecidable" microstates are colored in gray. Among the probability peaks showcased in Fig.\ref{Fig4}(e), the most prominent peaks associated with solution factors $(p,q)$ are $(101,111)$ and $(111,101)$, with probabilities of 58.4\% and 18.2\%, respectively, out of 5,003 non-gray events. Consequently, the solution to the factor pair in the problem of factoring $p \times q = 35$ is determined to be $(p,q) = (5,7)$ or $(7,5)$ based on the highest probability peaks observed in the experimental results in Fig.~\ref{Fig4}(e).

\section{Discussion} \label{SecV} 

It is worthwhile to discuss scaling issues related to the Rydberg-atom approach to the factorization problem. First, we will provide an estimation of the required number of atoms for encoding the integer factorization problem and then we will consider computational complexities associated with the presented reduction algorithm in the context of a Rydberg-atom experiment.

\textit{Atom resource estimation (upper bound).} The necessary number of atoms for the integer factorization of $n=p\times q$ in a Rydberg-atom experiment, is estimated as
\begin{equation}
N_{atom}=4.88N_C^{1.8}\approx 7.29 (\log_2 n)^{5.4}\label{eqn:Natom}
\end{equation}
in terms of $N_C$, the number of clauses~\cite{Jeong2023}, in the Boolean formula $\Psi$ designed for factoring the given integer $n$  (to be detailed below). 

\begin{figure*}[thbp]
  \centering
\includegraphics[width=0.98\textwidth]{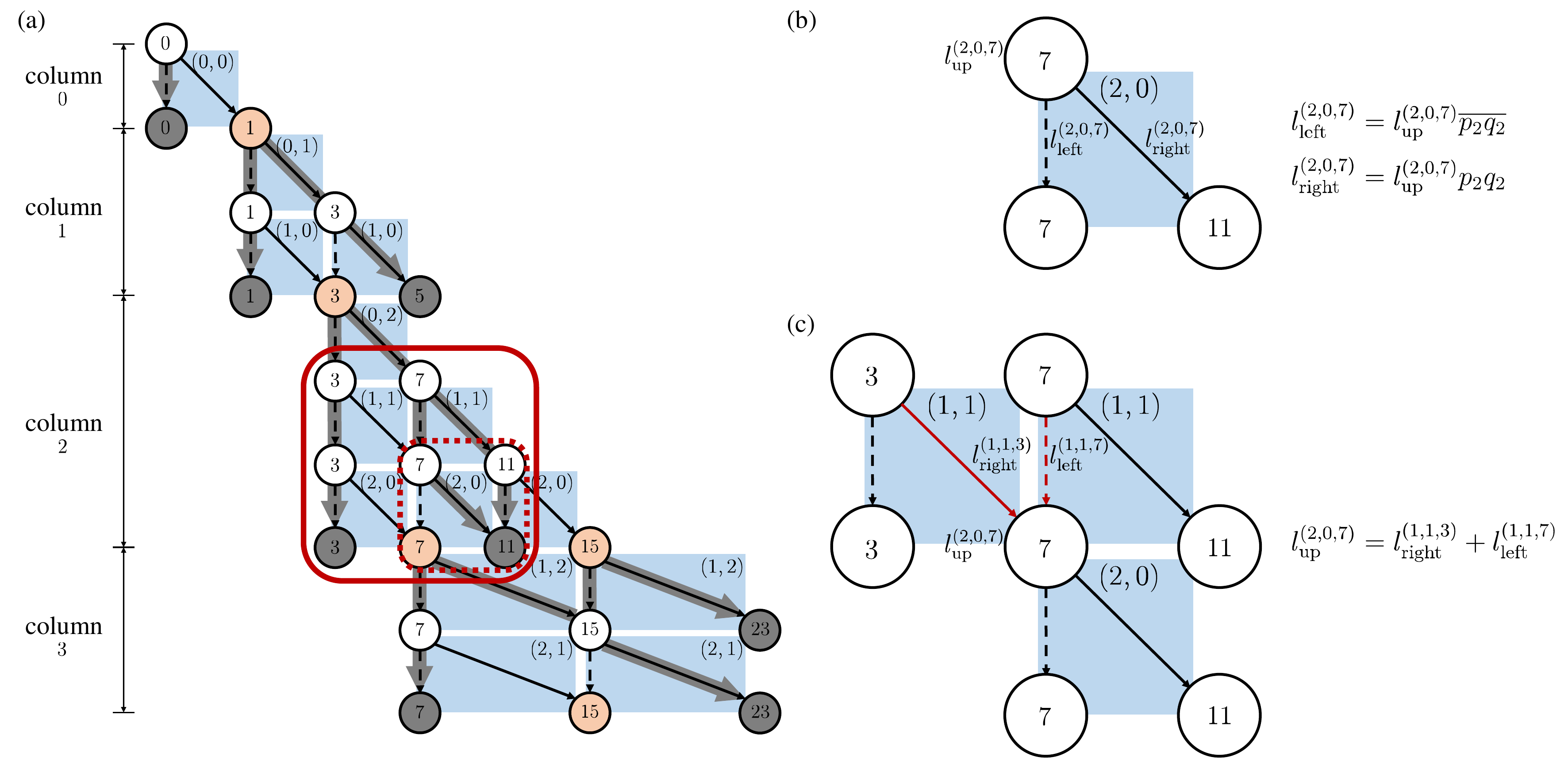}
\caption{(a) The binary decision diagram (BDD) designed for factoring $p\times q=15$. (b) A unit BDD responsible for constructing the 3-SAT formula $\Psi^{(2,0;7)}$. (c) Three unit BDDs collectively constructing the SAT formula $\Phi^{(2,0;7)}$. Refer to the text for a detailed discussion.}
\label{Fig5}
\end{figure*}

The determination of $N_C$ involves connecting each unit BDD corresponding to $f_{(i',j')}$ in Eq.~\eqref{eq:ij-term}, as illustrated in Fig.~\ref{Fig2}(b). For simplicity, we omit primes, mapping $i'\mapsto i$ and $j'\mapsto j$, hereafter. To approximate $N_C$, let us first compute the number of clauses corresponding to a single generic unit BDD. For example, in an exemplary BDD in Fig.~\ref{Fig5}(a) for factoring $p\times q=15$, there are 14 unit BDDs, each of which is represented with an initial running sum $v$ and parameters $p_i, q_j$, along with auxiliary variables $l_{\rm up}^{(i,j;v)}, l_{\rm left}^{(i,j;v)}, l_{\rm right}^{(i,j;v)} \in \{0,1\}$. A generic unit BDD shown in Fig.~\ref{Fig5}(b) corresponds to the assignment of these auxiliary variables for $i=2$, $j=0$, $v=7$, extracted from Fig.~\ref{Fig5}(a). The uppermost initial node of the unit BDD in Fig.~\ref{Fig5}(b) is assigned $l_{\rm up}^{(i,j;v)}=1$ if that node is ``passed-through'', and $l_{\rm up}^{(i,j;v)}=0$ otherwise. $l_{\rm left}^{(i,j;v)}$ and $l_{\rm right}^{(i,j;v)}$ are respectively assigned $1$ if the left bottom node and the right bottom node of the unit BDD is ``passed-through'', respectively, as given by
\begin{subequations}
\begin{eqnarray}
l_{\rm left}^{(i,j;v)} &=& l_{\rm up}^{(i,j;v)} \cdot \overline{p_i q_j},\\
l_{\rm right}^{(i,j;v)} &=& l_{\rm up}^{(i,j;v)} \cdot p_i q_j.
\end{eqnarray}  
\end{subequations}
These constraints among the auxiliary variables have to be satisfied and can be rewritten in a 3-SAT form as
\begin{eqnarray}
\Psi^{(i,j;v)}&=&(l_{\rm up}^{(i,j;v)}+\overline{l}_{\rm left}^{(i,j;v)})(\overline{p}_i+\overline{q}_j+\overline{l}_{\rm left}^{(i,j;v)})\nonumber \\
&&\cdot(p_i+\overline{l}_{\rm up}^{(i,j;v)}+l_{\rm left}^{(i,j;v)})(q_j+\overline{l}_{\rm up}^{(i,j;v)}+l_{\rm left}^{(i,j;v)})\nonumber \\
&&\cdot(l_{\rm right}^{(i,j;v)}+\overline{l}_{\rm up}^{(i,j;v)}+l_{\rm left}^{(i,j;v)})(\overline{l}_{\rm right}^{(i,j;v)}+l_{\rm up}^{(i,j;v)})\nonumber\\
&&\cdot(\overline{l}_{\rm right}^{(i,j;v)}+\overline{l}_{\rm left}^{(i,j;v)})=1.
\end{eqnarray}
The subsequent step involves establishing logical connections among the ``pass-through'' variables for various $(i,j;v)$ trios, such as those among $(i,j,v)=(1,1;3)$, $(1,1;7)$, and $(2,0;7)$ in Fig.~\ref{Fig5}(c). The initial top node of the unit BDD denoted by $(i,j)$ with a running sum $v$ on that node has two incoming edges from the top initial nodes of unit BDDs denoted by $(i,j)$ with running sums $v-2^{i+j}$ and $v$. So, a Boolean expression is established among the corresponding ``pass-through'' variables, $l_{\rm up}^{(i,j;v)}$, $l_{\rm right}^{(i-1,j+1;v-2^{i+j})}$, and $l_{\rm left}^{(i-1,j+1;v)}$, as follows:
\begin{equation}
l_{\rm up}^{(i,j;v)} = l_{\rm left}^{(i-1,j+1;v)}+l_{\rm right}^{(i-1,j+1;v-2^{i+j})}.
\end{equation}
These constraint for the BDD connections can be rewritten in a 3-SAT form as
\begin{eqnarray}
&& \Phi_{(i,j;v)}^{(i-1,j+1;v-2^{i+j});(i-1,j+1;v)} \nonumber \\
&&=(\overline{l}_{\rm up}^{(i,j;v)}+l_{\rm left}^{(i-1,j+1;v)}+l_{\rm right}^{(i-1,j+1;v-2^{i+j})}) \nonumber \\
&&\cdot (l_{\rm up}^{(i,j;v)}+\overline{l}_{\rm left}^{(i-1,j+1;v)}) \nonumber \\
&&\cdot (l_{\rm up}^{(i,j;v)}+\overline{l}_{\rm right}^{(i-1,j+1;v-2^{i+j})})=1.
\end{eqnarray}

Upon collecting all the unit-BDD 3-SAT formulas, $\Psi^{(i,j;v)}$, and their connections, $\Phi_{(i,j;v)}^{(i-1,j+1;v-2^{i+j});(i-1,j+1;v)}$, and logically multiplying them, we can obtain the total 3-SAT formula $\Psi_n$ for factoring the given integer $n$. The total number of clauses is then determined by counting the occurrences of the product, each generating $3+7=10$ clauses in the total 3-SAT formula. Since each unit BDD in Fig.~\ref{Fig2}(b) consists of three nodes, with two nodes overlapping among neighboring unit BDDs, each unit BDD effectively contributes two nodes to the total BDD. Consequently, the number of unit BDDs in the total BDD is obtained by dividing the total number of nodes in the total BDD by two. The number of nodes in the total BDD for factoring $n$, denoted as $B_n$, is determined by~\cite{Raddum2019}
\begin{equation}
2N_p^3-2N_p^2-2N_p+5 \lesssim B_n \lesssim 2N_p^3 -4N_p + 5,
\end{equation}
wherein we assume $p$ and $q$ to be $N_p=N_q = \log_2 n/2$-bit binary integers. Thus, the number of clauses in the total 3-SAT formula is, up to the leading order,
\begin{equation}
N_{C}=10\times(B_n /2)=10(\log_2 n/2)^3,
\end{equation}
which results in Eq.~\eqref{eqn:Natom}.

{\it Computational complexity.} The classical computational complexity involved in converting the integer factorization problem to a BDD is discussed following the methodology outlined in Ref.~\cite{Raddum2019}. The number of nodes in a total BDD is given by
\begin{equation} 
B_n \approx  2(\log_2 n/2)^3,
\end{equation}
when the BDD consists of $N_0 \equiv (\log_2 n /2)^3$ unit BDD blocks, each with an average 2 nodes, as depicted in Fig.~\ref{Fig2}(b). The number of time steps required to build each unit BDD of the total BDD for factoring the given integer $n$ and the memory space needed to arrange these unit BDD cells are respectively given by 
\begin{eqnarray}
\Delta N_{\mathrm{step}} &=& N_0, \label{eqn:time step requirement}\\
\Delta N_{\mathrm{memory}} &=& N_0. \label{eqn:memory requirement}
\end{eqnarray}
Hence, the construction of a BDD can be accomplished in polynomial time steps and memory space, efficiently using classical computation.

\section{Conclusion}\label{SecVI}
These Rydberg-atom experiments have taken on the task of addressing the integer factorization problem, with a particular focus on instances of $p\times q=6$, 15, and 35. The approach involves converting these instances into 3-SAT problems and subsequently mapping them onto Rydberg atom graphs. These graphs are then subjected to quasi-adiabatic quantum experiments, producing superpositions of microstates. These microstates are used to experimentally determine the integer factors $(p,q)$ that constitute $n=p\times q$. The proposed method estimates that the number of required atoms and classical computational resources for obtaining the Rydberg atom graph remain within polynomial orders of $\log_2 n$, suggesting the effectiveness of this encoding scheme.
Nonetheless, it is important to note that solving 3-SAT problems on a large scale using Rydberg atoms remains challenging, primarily due to the current limitations of imperfect quantum adiabatic processing hardware.

\section{Data availability} 
The experimental data set is archived in Ref.~\cite{ExpData2023} for further analysis.

\section{Acknowledgements}  \noindent
This research is supported by Samsung Science and Technology Foundation (SSTF-BA1301-52).
Louis Vignoli thanks Sélim Touati for useful discussions.

\begin{widetext}
\begin{appendix} 
\section{The Boolean expression for $p\times q=15$}

The binary decision diagram (BDD) in Fig.~\ref{Fig5}(a) is used to derive the SAT formula, $\Psi_{15}$ in Eq.~\ref{Psi15a}. The overall SAT formula for the integer factorization problem is expressed as:
\begin{equation}
\Psi_{15}=\Psi_{15,0}  \Psi_{15,1}  \Psi_{15,2}  \Psi_{15,3},
\end{equation}
where $\Psi_{15,0}$, $\Psi_{15,1}$, $\Psi_{15,2}$, and $\Psi_{15,0}$ correspond to columns 0, 1, 2, and 3 of the BDD, representing bitwise relations for $p=(p_2 p_1 p_0)_2 \times (q_2 q_1 q_0)_2 = (1111)_2$, respectively. In Fig.~\ref{Fig5}(a), the orange-colored circles indicate nodes that need to be "passed-through" for successful factorization of $n$, while the gray-colored filled circles are not to be "passed-through."

The unsuccessful path in column 0 corresponds to $\overline{p_0q_0}=1$. Consequently, the Boolean subformula for column 0 is $\neg\Psi_{15,0}=\overline{p_0q_0}$
resulting in the b=Boolean subformula for column 0 in conjunctive normal form:\begin{equation}
\Psi_{15,0}=p_0q_0
\end{equation}
Similarly, for columns 1 and 2, $\Psi_{15,1}$ and $\Psi_{15,2}$ are obtained by aggregating failed paths in the respective columns:
\begin{eqnarray}
\Psi_{15,1}&=&\neg(\overline{p_0q_1}\cdot\overline{p_1q_0}+p_0q_1\cdot p_1q_0)\nonumber\\
&=&(p_0+p_1)(p_0+q_0)(q_1+p_1)(q_1+q_0)({\overline{p}}_0+{\overline{q}}_1+{\overline{p}}_1+{\overline{q}}_0), \\
\Psi_{15,2}&=&\neg(\overline{p_0q_2}\cdot\overline{p_1q_1}\cdot\overline{p_2q_0}+\overline{p_0q_2}\cdot p_1q_1\cdot p_2q_0 +p_0q_2\cdot\overline{p_1q_1}\cdot p_2q_0+p_0q_2\cdot p_1q_1\cdot\overline{p_2q_0})\nonumber\\
&=&(p_0q_2+p_1q_1+p_2q_0)(p_0q_2+\overline{p_1q_1}+\overline{p_2q_0})
(\overline{p_0q_2}+p_1q_1+\overline{p_2q_0})(\overline{p_0q_2}+\overline{p_1q_1}+p_2q_0).
\end{eqnarray}
For column 3, paths starting at either the orange-colored `7' node or `15' node on the same horizontal position in Fig.~\ref{Fig5}(a) are considered. Failed paths $\overline{p_1q_2}\cdot\overline{p_2q_1}=1$, $p_1q_2\cdot p_2q_1=1$ starting at the orange-colored `7' node, and failed paths $p_1q_2=1$, $\overline{p_1q_2}\cdot p_2q_1=1$ starting at the `15' node in column 3 are identified. To distinguish between these two distinct failed paths, we introduce auxiliary variables $l_A, l_B \in {0,1}$, with values set to 1 only if the `7' (`15') node is "passed-through." The Boolean subformula corresponding to column 3 is then expressed as:
\begin{eqnarray}
\Psi_{15,3}&=&\{l_A\rightarrow\neg(\overline{p_1q_2}\cdot\overline{p_2q_1}+p_1q_2\cdot p_2q_1)\}\{l_B\rightarrow\neg(p_1q_2+\overline{p_1q_2}\cdot p_2q_1)\}\nonumber\\
&=&\{l_A\rightarrow(p_1+p_2)(p_1+q_1)(q_2+p_2)(q_2+q_1)({\overline{p}}_1+{\overline{q}}_2+{\overline{p}}_2+{\overline{q}}_1)\}\nonumber \\
&&\cdot\{l_B\rightarrow({\overline{p}}_1+{\overline{q}}_2)(p_1+{\overline{p}}_2+{\overline{q}}_1)(q_2+{\overline{p}}_2+{\overline{q}}_1)\}\nonumber\\
&=&\{\overline{l}_A+(p_1+p_2)(p_1+q_1)(q_2+p_2)(q_2+q_1)({\overline{p}}_1+{\overline{q}}_2+{\overline{p}}_2+{\overline{q}}_1)\}\cdot\{{\overline{l}}_B+({\overline{p}}_1+{\overline{q}}_2)\nonumber\\
&&\cdot(p_1+{\overline{p}}_2+{\overline{q}}_1)(q_2+{\overline{p}}_2+{\overline{q}}_1)\},
\end{eqnarray}
where $l_A$ is expressed in terms of the sum of non-failed paths in column 2:  
$\overline{p_0q_2}\cdot p_1q_1\cdot p_2q_0=1$, $\overline{p_0q_2}\cdot p_1q_1\cdot\overline{p_2q_0}=1$, and $p_0q_2\cdot\overline{p_1q_1}\cdot\overline{p_2q_0}=1$, which lead to the orange-colored `7' node:
\begin{equation}
l_A=\overline{p_0q_2}\cdot p_1q_1\cdot p_2q_0+\overline{p_0q_2}\cdot p_1q_1\cdot\overline{p_2q_0}+p_0q_2\cdot\overline{p_1q_1}\cdot\overline{p_2q_0}
\end{equation}
and $l_B$ is expressed in terms of the non-failed path $p_0 q_2 \cdot p_1 q_1 \cdot p_2 q_0=1$ in column 2, which leads to the `15' node:
\begin{equation}
l_B=p_0q_2\cdot p_1q_1\cdot p_2q_0.
\end{equation}
Substituting $l_A, l_B$ into $\Psi_{15,3}$ yields a Boolean equation in terms only of $p,q$. 

Simplifying $\Psi_{15}$ involves utilizing the fact that $\Psi_{15,0}=p_0 q_0 = 1$, as single variable clauses $p_0$ and $q_0$ trivially result in values of $p_0$ and $q_0$ as 1. Substituting $p_0=q_0=1$, the expressions for $\Psi_{15,1}$, $\Psi_{15,2}$, $l_A$, and $l_B$ become:
\begin{eqnarray}
\Psi_{15,1}&=&(p_1+q_1)({\overline{p}}_1+{\overline{q}}_1) \\
\Psi_{15,2}&=&(q_2+p_1q_1+p_2)(q_2+{\overline{p}}_1+{\overline{q}}_1+{\overline{p}}_2) ({\overline{q}}_2+p_1q_1+{\overline{p}}_2)({\overline{q}}_2+{\overline{p}}_1+{\overline{q}}_1+p_2)\nonumber\\
&=&(q_2+p_1+p_2)(q_2+q_1+p_2)(q_2+{\overline{p}}_1+{\overline{q}}_1+{\overline{p}}_2)({\overline{q}}_2+p_1+{\overline{p}}_2)({\overline{q}}_2+q_1+{\overline{p}}_2)({\overline{q}}_2+{\overline{p}}_1+{\overline{q}}_1+p_2) \nonumber \\
l_A&=&\overline{q_2}\cdot p_1q_1\cdot p_2+\overline{q_2}\cdot p_1q_1\cdot\overline{p_2}+q_2\cdot\overline{p_1q_1}\cdot\overline{p_2}=\overline{q_2}p_1q_1p_2+\overline{q_2}p_1q_1\overline{p_2}+q_2{\overline{p}}_1{\overline{p}}_2+q_2{\overline{q}}_1{\overline{p}}_2 \\
l_B&=&q_2p_1q_1p_2.
\end{eqnarray}
Substituting $l_A$ and $l_B$ into $\Psi_{15,3}$ yields:
\begin{eqnarray}
\Psi_{15,3}&=&\{(q_2+p_1+{\overline{p}}_2)(q_2+q_1+{\overline{p}}_2)(q_2+{\overline{p}}_1+{\overline{q}}_1+p_2)({\overline{q}}_2+p_1+p_2)({\overline{q}}_2+q_1+p_2)+(p_1+p_2)(p_1+q_1)\nonumber\\
&&\cdot(q_2+p_2)(q_2+q_1)({\overline{p}}_1+{\overline{q}}_2+{\overline{p}}_2+{\overline{q}}_1)\}\{({\overline{q}}_2+{\overline{p}}_1+{\overline{q}}_1+{\overline{p}}_2)+({\overline{p}}_1+{\overline{q}}_2)(p_1+{\overline{p}}_2+{\overline{q}}_1)\nonumber\\
&&\cdot(q_2+{\overline{p}}_2+{\overline{q}}_1)\} \nonumber \\
&=& ({\overline{q}}_2+p_1+p_2)(q_2+q_1+{\overline{p}}_2)({\overline{q}}_2+{\overline{p}}_1+{\overline{q}}_1+{\overline{p}}_2).
\end{eqnarray}
Multiplying every subformula for each of the columns 0, 1, 2, and 3 results in the following form:
\begin{eqnarray}
\Psi_{15} 
&=&p_0q_0\times (p_1+q_1)({\overline{p}}_1+{\overline{q}}_1)(q_2+p_1+p_2)(q_2+q_1+p_2)(q_2+{\overline{p}}_1+{\overline{q}}_1+{\overline{p}}_2)\nonumber\\
&&\cdot ({\overline{q}}_2+p_1+{\overline{p}}_2)({\overline{q}}_2+q_1+{\overline{p}}_2)({\overline{q}}_2+{\overline{p}}_1+{\overline{q}}_1+p_2)({\overline{q}}_2+p_1+p_2)(q_2+q_1+{\overline{p}}_2)\cdot({\overline{q}}_2+{\overline{p}}_1+{\overline{q}}_1+{\overline{p}}_2).
\end{eqnarray}
After simplification using Boolean identities $A(A+B)=A$ and $(A+B)(A+\overline{B})=A$, the SAT formula corresponding to the problem of factoring $p\times q =15$ is obtained as:
\begin{eqnarray}
\Psi_{15}&=&(p_1+p_2)(q_1+q_2)({\overline{p}}_1+{\overline{q}}_1)(p_1+q_1)
(p_1+{\overline{p}}_2+{\overline{q}}_2)(q_1+{\overline{q}}_2+{\overline{p}}_2)p_0q_0.
\end{eqnarray}

\end{appendix}

\end{widetext}

\end{document}